\documentclass[12pt]{article}
\usepackage{amssymb}
\usepackage{amsmath,,calrsfs}
\usepackage[dvips]{epsfig}

\newcommand{\be}{\begin{equation}}
\newcommand{\ee}{\end{equation}}
\def\bea{\begin{eqnarray}}
\def\eea{\end{eqnarray}}

\newcommand{\bn}{\begin{eqnarray}}
\newcommand{\en}{\end{eqnarray}}

\newcommand{\p}{\partial}

\newcommand{\no}{\noindent}

\newcommand{\s}{\,\,\,\,}
\def\bea{\begin{eqnarray}}
\def\eea{\end{eqnarray}}

\newcommand{\beq}{\begin{eqnarray}}
\newcommand{\eeq}{\end{eqnarray}}
\begin{document}

\title{\textbf{Soldering spin-3/2 fermions in $D=2+1$ }}
\author{E. L. Mendon\c ca \footnote{eliasleite@feg.unesp.br}, D. S. Lima\footnote{diegos.lima88@hotmail.com }, A. L. R. dos Santos \footnote{alessandroribeiros@yahoo.com.br} \\
\textit{{UNESP - Campus de Guaratinguet\'a - DFQ} }\\
\textit{{Av. Dr. Ariberto Pereira da Cunha, 333} }\\
\textit{{CEP 12516-410 - Guaratinguet\'a - SP - Brazil.} }\\
}
\date{\today}
\maketitle

\begin{abstract}
The soldering procedure has been for the first time generalized to the case of spin-3/2 fermionic theories. We have demonstrated that the fermionic part of the so called ``New Topologically Massive Supergravity'' theory, which is of third order in derivatives, can be soldered in order to obtain a fourth order dublet model analogue to the linearized version of the New Massive Gravity theory, while the soldering of two second order self-dual models give us a theory similar to the linearized version of the Einstein-Hilbert-Fierz-Pauli theory.
\end{abstract}
\newpage

\section{Introduction}

In $D=2+1$ dimensions self-dual models describes massive particles with spin $+s$ or $-s$. Depending on the spin, we have a different number of self-dual descriptions. In the spin-1 case for example we have two self-dual descriptions, while in the spin-2 case we have four. The models differ each other by an order in derivatives and by gauge invariance under a given symmetry or a set of them. 

We have studied in \cite{nges32} the connection among three massive spin-3/2  self-dual versions of first, second and third order in derivatives through symmetry arguments. We have also considered an analogue version of the Proca (or Fierz-Pauli) theory which is second order in derivatives and which by means of the Fierz-Pauli conditions describes correctly two massive propagating modes of helicities $+3/2$ and $-3/2$. This doublet model \footnote{All along the work we use the terminology of singlet models meaning that the models carries only a unique helicity $+3/2$ or $-3/2$, while the dublet models carries both of them together.} is by its turn connected to a fourth order doublet model which we believe is free of ghosts. We have argued that, due the similarities, the fourth order model could be interpreted as the spin-3/2 version of the linearized New Massive Gravity ($NMG$) \cite{BHT}. However, it is  unclear so far if it is possible to obtain the doublet models from the self-dual ones as well as the first order self-dual model from the decomposition of the second order dublet model.  That is what we are going to address in the following lines using the soldering approach, which we briefly describe in the next paragraphs.

The soldering approach is only one of the several tools we have to investigate dualities. The technique was first introduced by Stone many years ago \cite{stone}, and since then it has been used on the study of dual aspects of physical systems with many applications in different contexts as well as in different dimensions. Electromagnetic dualities can be explored through the soldering procedure providing new point of views of the same physics \cite{3clovis}. The quantization of strings and also other theoretical models encouraged the study of chiral bosons in the nineties. In this context the authors in \cite{DAE} (See also references therein) have used the soldering formalism to combine the opposite chiralities of the Schwinger model. As a consequence they have noticed that the formalism could be generalized leading us either to the vector or to the axial models depending on a constant parameter $\alpha=\pm 1$. A brief and interesting note \cite{Ghosh} has showed us that even in the presence of non-comutative field theories the soldering formalism can be applied. Although the several generalizations and applications of the technique we do not know an example of the soldering procedure applied to the case of fermionic fields.

In \cite{ad1} the authors have noticed that in the spin-1 case the first order self-dual model can be understood as the ``square root'' of the Proca theory (which carries a dublet of spins $\pm 1$), in such a way that it is equivalent to the direct sum of  first order self-dual models with opposite helicities. We can notice that once the first-order self-dual model is equivalent to the Maxwell-Chern-Simons ($MCS$) model (the second order description) one can expect that the direct sum of two $MCS$ models will also produce the Proca theory, but that is not true. To obtain the Proca theory from the ``sum'' of $MCS$ models we need another approach which is the soldering procedure. We can understand that since the first-order self-dual description does not enjoy of the gauge invariance, so their direct sum results in the non-gauge invariant Proca theory, on the other hand since the $MCS$ models are gauge invariant the Proca theory is achieved only through the addition of an interference term in the sum, which is provided by means of the soldering procedure. One can generalize the soldering procedure in order to adapting it to the case of different  masses ($m_-$ and $m_+$ for instance), in this case by soldering two $MCS$ models we obtain a generalized version of the Proca theory containing an additional Chern-Simons term proportional to the difference of  masses ($m_+-m_-$) such that once we have $m_+=m_-$ the Proca theory is recovered.

The scenario becomes interesting when we have other higher order self-dual descriptions as it is the case when we are dealing with spin-2 particles \cite{nsd,Andringa}. We have noticed that, like in the spin-1 case the Fierz-Pauli theory (which carries a dublet of spins $\pm 2$) can also be decomposed in a pair of first-order self-dual models. Likewise, the soldering of two second-order self-dual models becomes the Fierz-Pauli theory if we have equal masses and an additional Chern-Simons-like term appears if we do with different masses. Surprisingly the procedure can be extended to the soldering of third and also fourth order self-dual models, the result in both cases are the same, the linearized version of the $NMG$ theory \cite{parity}. Generalizing the procedure for different masses an additional linearized topological Chern-Simons term becomes part of the result.

Some discussion about the procedure we have summarized above and its features is well detailed in \cite{parity, gss2}. Given the versatility of this technique one can think in some more generalizations. We have tried the extension of this method to the case of spin-3 self-dual models, but unfortunately the presence of auxiliary fields can not be bypassed, and only the spin-3 sector of the lagrangians can be soldered correctly. However is an original attempt to try the soldering of fermionic models which do not demand the presence of auxiliary fields. This is precisely the case of the massive spin-3/2 theories.

In this work we use the massive spin-3/2 self-dual descriptions in $D=2+1$ as a laboratory to the soldering procedure in order to construct dublet models from singlet ones. The dublet models we have obtained here can be interpreted as analogue versions of the $MCS$-Proca model or in other case as the linearized version of the $NMG$ theory. 

\section{Obtaining singlets from dublets}

Let us consider the dublet model for spin-3/2 particles in $ D=2+1$

\begin{equation}
S_{d}^{(2)} = \int d^3x \left[ -\frac{1}{4}\bar{f}^{\mu}(\psi)\gamma_{\nu}\gamma_{\mu}f^{\nu}(\psi) + \frac{m^2}{2}\epsilon^{\mu \nu \beta}\bar{\psi}_{\mu}\gamma_{\nu}\psi_{\beta}\right]\label{dublet}
\end{equation}

First of all we need to set up some points on the notations and other details. The fields are two component Majorana vector-spinors, and the greek indices corresponds to the space-time while the metric is mostly plus $(-,+,+)$. The spinorial indices has been suppressed for sake of simplicity, and the gamma matrices are indeed the Pauli matrices in agreement with the notation adopted in \cite{djt}. We have been working with doublet and singlet models which are connected each other. To avoid any kind confusion we adopt the greek letters $\psi$ and $\chi$ to describe the fermionic fields of dublet actions, and capital latin letters to describe the fermionic fields of singlet models. The actions will be followed by a superscript $(n)$ with $n$ indicating its order in derivatives and by a subscript indicating the helicity. The subscript $d$ refers to the dublet of helicities $\pm 3/2$. Besides, we have used the same notation of \cite{deserkay, deser3/2} where $\bar{f}^{\mu}(\psi) = \epsilon^{\mu \alpha \beta}\partial_{\alpha}\bar{\psi}_{\beta}$ and $
f^{\mu}(\psi) = \epsilon^{\mu \alpha \beta}\partial_{\alpha}{\psi}_{\beta}$.

As we have demonstrated before in \cite{nges32} the action above carries two degrees of freedom corresponding to the massive modes of helicities $+3/2$ and $-3/2$. In order to decompose the doublet model in two singlet self-dual models we need to lower the order in derivatives of the second order term, which can be made with the help of auxiliary fields $\chi_{\mu}$ and $\bar{\chi_{\mu}}$ in the following way:

\begin{equation}
S_{d}^{(1)} = \int d^3x \left[ -\frac{m}{2} \bar{\chi}_{\mu}f^{\mu}(\psi) - \frac{m}{2} \bar{f}^{\mu}(\psi)\chi_{\mu} + \frac{m^2}{2}\epsilon^{\mu \nu \alpha}\bar{\chi}_{\mu}\gamma_{\nu}\chi_{\alpha} + \frac{m^2}{2}\epsilon^{\mu \nu \beta}\bar{\psi}_{\mu}\gamma_{\nu}\psi_{\beta}\right]\label{two}.  
\end{equation}

By taking the equations of motion for the auxiliary fields in (\ref{two}) and plugging back the results in the action is straightforward to see that we recover (\ref{dublet}). Then one can proceed with a rotation of the fields in order to decouple the action (\ref{two}). The rotation is given by:

\begin{equation}
\bar{\chi}_{\mu} = (\bar{A}_{\mu} + \bar{B}_{\mu}) \qquad;\qquad \chi_{\mu} =(A_{\mu} + B_{\mu})\label{rot1},
\end{equation}

\begin{equation}
\bar{\psi}_{\mu} = (\bar{B}_{\mu} - \bar{A}_{\mu} ) \qquad;\qquad \psi_{\mu} =(B_{\mu} - A_{\mu})\label{rot2},
\end{equation}

\noindent substituting (\ref{rot1}) and (\ref{rot2}) in  (\ref{two}), we have the completely decoupled pair of first order self-dual models representing the different helicities $+3/2$ and $-3/2$:

\begin{eqnarray}
S_d &=& \int d^3x \left[-m\bar{A}_{\mu}f^{\mu}(A) - m^2\epsilon^{\mu \nu \alpha}\bar{A}_{\mu}\gamma_{\nu}A_{\alpha}	+ m\bar{B}_{\mu}f^{\mu}(B) - m^2\epsilon^{\mu \nu \alpha}\bar{B}_{\mu}\gamma_{\nu}B_{\alpha}\right]\\
&=& S_{{-3/2}}^{(1)}[A] + S_{{+3/2}}^{(1)}[B], \label{OBTAD}
\end{eqnarray}

\no which is in completely analogy with the spin-1 and spin-2 cases. In a certain way we have taken the square-root of the doublet model separating the helicities apart. In the next section the soldering will be used to obtain the doublet model from the self-dual ones. The normal coordinates given by (\ref{rot1}) and (\ref{rot2}) have been chosen in such a way that the result given by (\ref{OBTAD}) is analogue to the spin-1 and spin-2 cases whereas the Chern-Simons term is preceded by a mass parameter which determines the helicity.

 \section{Soldering second order self-dual models}
In the previous case,  we known that the model (1) describes in $D=2+1$ dimensions a parity doublet of helicities $+3/2$ and $ - 3/2$ which is the same particle content of the two first order self-dual models of opposite helicities. Since both of them have no local symmetries one has been able to relate them through a trivial rotation. However, regarding its dual theories, a pair of second and third order self-dual models of opposite helicities, it is not so easy to identify it with the dublet theory due to the presence of local symmetries in such theories. One needs an extra “interference term” between the opposite helicities in order to comply with the local symmetries. This extra term can be produced by the soldering formalism. The physical idea of fusing two fields representing complementary aspects of some symmetry into one specific combination of fields is the core of the present techique we will use in this section.

Here we start with the second order self-dual models describing the helicities $+3/2$ or $-3/2$ in $D=2+1$ dimensions. The authors in \cite{deserkay} have showed that the second order model is equivalent by means of a master action to the first order self-dual one. On the other hand we have demonstrated that this model can be systematically generated through the Noether Gauge Embedment approach starting with the non gauge invariant first order model. 

Let us consider the pair of actions.

\begin{equation}
S^{(2)}_{+3/2}[A]=\int d^3x \left[ -\frac{1}{4}\bar{f}^{\mu}(A)\gamma_{\nu}\gamma_{\mu}f^{\nu}(A) + \frac{m_+}{2}\bar{A}_{\mu}f^{\mu}(A)\right], \label{spin+}
\end{equation}

\begin{equation}
S^{(2)}_{-3/2}[B]=\int d^3x \left[ -\frac{1}{4}\bar{f}^{\mu}(B)\gamma_{\nu}\gamma_{\mu}f^{\nu}(B) - \frac{m_-}{2}\bar{B}_{\mu}f^{\mu}(B) \label{spin-}\right].
\end{equation}
We have defined the sign of the mass terms in order to indicate the correspondent helicity, in such a way that (\ref{spin+}) descibes a massive spin $+3/2$ particle with mass $m_+$, while (\ref{spin-}) describes a massive spin $-3/2$ particle with mass $m_-$. One can also verify that the hole action (\ref{spin+}) (or (\ref{spin-})) is gauge invariant under  transformations of the type $\delta A_{\mu}=\p_{\mu}\xi$ and $\delta\bar{A}_{\mu}=\p_{\mu}\bar{\xi}$ with $\xi$ and $\bar{\xi}$ spinorial parameters depending on the space-time coordinates. After an integration by parts we can also verify that they are invariant under the global shifts $\delta A_{\mu}=\Lambda_{\mu}$ and $\delta\bar{A}_{\mu}=\bar{\Lambda}_{\mu}$. The soldering procedure consists indeed in to lift the global shifts to local symmetries tying the fields $A_{\mu}$ and $B_{\mu}$ (as well as their adjuncts) together by imposing that their variations are proportional to each other, in such a way that we have:

\begin{equation}
\delta A_{\mu}=\Lambda_{\mu} \qquad;\qquad \delta\bar{A}_{\mu} = \bar{\Lambda}_{\mu} \label{simetria1}
\end{equation}

\begin{equation}
\delta B_{\mu}=\alpha\Lambda_{\mu} \qquad;\qquad \delta\bar{B}_{\mu} = \alpha\bar{\Lambda}_{\mu},\label{simetria2}
\end{equation}

\no where we have introduced the proportionality constant $\alpha$ so far arbitrary. \\
\indent  Taking the variation of both actions and then summing the results we have:

\begin{equation}
\begin{split}
\delta\left(S^{(2)}_{+3/2}[A] + S^{(2)}_{-3/2}[B]\right) = \int d^{3}x \s\left\lbrace \bar{f}^{\mu}(\Lambda)\left[ -\frac{1}{4}\gamma_{\nu}\gamma_{\mu}{f}^{\nu}(A + \alpha B) + \frac{1}{2}(m_+A_{\mu} - m_-\alpha B_{\mu})\right]\right.\\ \left.
+\left[  -\frac{1}{4}\bar{f}^{\mu}(A + \alpha B)\gamma_{\nu}\gamma_{\mu} +  \frac{1}{2}(m_+\bar{A}_{\nu} - m_-\alpha\bar{B}_{\nu})\right]f^{\nu}(\Lambda)\right\rbrace. \label{corrente}
\end{split}
\end{equation}

\no In (\ref{corrente}) it useful to define the soldering currents $J_{\mu}$ and $\bar{J}_{\mu}$ by the expressions between brackets multiplying $\bar{f}^{\mu}(\Lambda)$ and $f^{\nu}(\Lambda)$ respectively, in such a way that we get the much compact expression:

\begin{equation}
\delta\left(S^{(2)}_{+3/2}[A] + S^{(2)}_{-3/2}[B]\right) = \int d^{3}x  \s \left[\bar{f}^{\mu}(\Lambda)J_{\mu} + \bar{J}_{\mu}f^{\mu}(\Lambda)\right]\label{corrente2}. 
\end{equation}

\noindent It is obvious then, that the simple sum of the actions do not give us an invariant theory under the set of transformations given by (\ref{simetria1}) and (\ref{simetria2}). An interference term must be included in order to get that. This can be done by considering auxiliary fields, which we will call $H_{\mu}$ and $\bar{H}_{\mu}$, such that their variations will exactly cancel the result obtained in (\ref{corrente2}). Then we have:

\begin{equation} 
\delta\bar{H}^{\mu}\equiv \bar{f}^{\mu}(\Lambda) \quad ; \quad \delta H^{\mu} \equiv f^{\mu}(\Lambda). 
\end{equation}
\no In the presence of the auxiliary field one can reorganize the variation in (\ref{corrente2}) such that:

\begin{equation}
\delta\left[S^{(2)}_{+3/2}[A] + S^{(2)}_{-3/2}[B] - \int d^{3}x (\bar{H}^{\mu}J_{\mu} + \bar{J}_{\mu}H^{\mu})\right] = -\int d^{3}x \s \left(\bar{H}^{\mu}\delta J_{\mu} + \delta \bar{J}_{\mu}H^{\mu}\right)\label{deltaj}.
\end{equation}

\no Where the variation of the currents are given by:

\begin{equation}
\delta J_{\mu} = -\frac{(1+\alpha^{2})}{4}\gamma_{\nu}\gamma_{\mu}\delta H^{\nu} + \frac{1}{2}(m_+ - m_-\alpha^{2})\Lambda_{\mu}\label{ble1},
\end{equation}

\begin{equation}
\delta \bar{J}_{\mu} = -\frac{(1+\alpha^{2})}{4}\delta \bar{H}^{\nu}\gamma_{\mu}\gamma_{\nu} + \frac{1}{2}(m_+ - m_-\alpha^{2})\Lambda_{\mu},\label{ble2}
\end{equation}

\no and we notice that the last terms in (\ref{ble1}) and (\ref{ble2}) can not be written in terms of $\delta H_{\mu}$ and $\delta \bar{H}_{\mu}$ respectively. But, since the parameter $\alpha$ is still arbitrary we choose it in order to get rid of these terms by making $\alpha^{2}={m_+}/{m_-}$. Then, substituting back (\ref{ble1}) and (\ref{ble2}) in (\ref{deltaj}), we automatically define the invariant soldered action, given by:

\begin{equation}
S_S= S^{(2)}_{+3/2}[A] + S^{(2)}_{-3/2}[B] - \int d^{3}x \left[\bar{H}^{\mu}J_{\mu} + \bar{J}_{\mu}H^{\mu} + \frac{(1+\alpha^{2})}{4}\bar{H}^{\mu}\gamma_{\nu}\gamma_{\mu}H^{\nu}\right].
\end{equation}

However, the auxiliary fields must be eliminated, which can be made through their equations of motion. After some algebra we end up with:

\begin{equation}
S_{S}^{(2)} = S^{(2)}_{+3/2}[A] + S^{(2)}_{-3/2}[B] + \frac{2}{(1 + \alpha^{2})}\int d^{3}x\s \left[\bar{J}_{\mu}J^{\mu} - \bar{J}_{\mu}\gamma^{\mu}\gamma^{\nu}J_{\nu}\right]. \label{JJ} 
\end{equation}

\no The reader can be spared from several tedious calculations, which consists of replacing the currents $\bar{J}_{\mu}$ and $J_{\mu}$ and the actions $S^{(2)}_{+3/2}[A]$ and $ S^{(2)}_{-3/2}[B]$ in (\ref{JJ}) to obtain an invariant action forced to depend only upon the exact combinations $\psi_{\mu}=\alpha A_{\mu}-B_{\mu}$ and $\bar{\psi}_{\mu}=\alpha \bar{A}_{\mu}-\bar{B}_{\mu}$, both of them invariant under the set of transformations (\ref{simetria1}) and (\ref{simetria2}). The final result can be written as:

\begin{align}
\begin{aligned}
S^{(2)}_d[\psi] = \frac{1}{(1+\alpha^2)}\int d^3x\, \left[ -\frac{1}{4}\bar{f}^{\mu}(\psi)\gamma_{\nu}\gamma_{\mu}f^{\nu}(\psi) + \frac{m_+m_-}{2}\epsilon^{\mu \nu \beta}\bar{\psi}_{\mu}\gamma_{\nu}\psi_{\beta}+ \frac{(m_+ - m_-)}{2}\bar{\psi}^{\mu}f_{\mu}(\psi)\right]
\end{aligned} \label{final}
\end{align}

Notice that we have used again the greek leter $\psi$ indicating the field, and stressed with the subscript $d$ that the action we have obtained describe a dublet of spins $+3/2$ and $-3/2$ with different masses $m_+$ and $m_-$. Besides, we have to say that once we consider the field $\psi_{\mu}$ as an independent arbitrary field (forgetting about the fact it is a invariant combination of $A$ and $B$) the action (\ref{final}) is obviously non gauge invariant under the transformations $\delta \psi_{\mu}=\p_{\mu}\xi$ and $\delta \bar{\psi}_{\mu}=\p_{\mu}\bar{\xi}$ due to the presence of the mass term. In some sense, this emphasizes that the soldering procedure is not a simple sum of the actions, otherwise the sum of the gauge invariant actions would give us a gauge invariant new action. The interference term proportional to $J^2$ in (\ref{JJ}) has introduced the gauge symmetry breaking.

On the result we have obtained in (\ref{final}), it is completely analogue to the cases of spins $1$ and $2$. Comparing with the spin-1 case one would say that we have a kind of Maxwell-Chern-Simons-Proca theory if the masses are different. On the other hand if $m_+=m_-$ we have a kind of Maxwell-Proca model. With equal masses, the model is exactly the same from where we have obtained the first order self-dual models (\ref{dublet}) by lowering the order and rotating the fields as we have done in the previous section. In that case, both of them, the dublet model and the first order self-dual models are non-gauge invariant, and that is why there is no sense on applying the soldering approach.

\section{Soldering third order self-dual models}

As we have said in the introduction there are three self-dual descriptions for massive spin-3/2 particles in $D=2+1$ dimensions. Here we have verified that the first-order one can be obtained from the second order dublet model given by (\ref{dublet}). Besides, by soldering two second order models the very same dublet model is recovered. The emergent question, is about the soldering of the third order self-dual descriptions. We have observed that, some thing very interesting happens in the case of spin-2 particles. In \cite{parity} we have demonstrated that the linearized version of the New Massive Gravity models can be understood as the soldering of two third order self-dual models, which in that case corresponds exactly to the linearized truncation of the topologically massive gravity. It is obvious that, the gravitational appeal becomes that calculation more interesting than the others, but since our third order model is indeed the fermionic part of the topologically massive supergravity theory of \cite{Andringa}, one could speculate about the fermionic part of a super-NMG theory wich would be a fourth order dublet model. To find this model we suggest the soldering of the following actions:

\begin{equation}
S^{(3)}_{+3/2}[A]=\int d^3x\, \left[\bar{f}^{\mu}(A)\gamma_{\nu}\gamma_{\mu}f^{\nu}(A) + \frac{1}{2m_+}\epsilon^{\alpha \lambda \beta}\bar{f}^{\mu}(A)\gamma_{\alpha}\gamma_{\mu}\gamma_{\nu}\gamma_{\beta}\partial_{\lambda}f^{\nu}(A)\right], \label{act3+}
\end{equation}

\begin{equation}
S^{(3)}_{-3/2}[B]=\int d^3x\, \left[\bar{f}^{\mu}(B)\gamma_{\nu}\gamma_{\mu}f^{\nu}(B) - \frac{1}{2m_-}\epsilon^{\alpha \lambda \beta}\bar{f}^{\mu}(B)\gamma_{\alpha}\gamma_{\mu}\gamma_{\nu}\gamma_{\beta}\partial_{\lambda}f^{\nu}(B)\right].\label{act3-}
\end{equation}

From now on there will be a proliferation of gamma matrices and lots of indices will be needed, to avoid any confusion we think that it is a good idea to define some shorthand notation by making $
\epsilon^{\alpha \lambda \beta}\partial_{\lambda} = E^{\alpha \beta}$ and $ \gamma_{\alpha}\gamma_{\mu}\gamma_{\nu}\gamma_{\beta}=\Delta_{\alpha \mu \nu \beta}$. As before we would like to tie the fields $A_{\mu}$ and $B_{\mu}$ (as well as their adjuncts) by considering that their variations are proportional each other, which can be done by means of the $\alpha$ arbitrary constant:

\begin{equation}
\delta A_{\mu} = \Lambda_{\mu}  \qquad;\qquad  \delta\bar{A}_{\mu} = \bar{\Lambda}_{\mu}\label{sym1},
\end{equation}

\begin{equation}
\delta B_{\mu} = \alpha\Lambda_{\mu}  \qquad;\qquad  \delta\bar{B}_{\mu} = \alpha\bar{\Lambda}_{\mu}\label{sym2},
\end{equation}

\noindent The symmetries (\ref{sym1}) and (\ref{sym2}) are global ones, and the soldering approach will lift it to local symmetries. It is straightforward to see that the actions (\ref{act3+}) and (\ref{act3-}) are invariant under these rigid transformations. But by this time, we could also check that they are invariant under a larger set of gauge transformations, which are: $\delta A_{\mu}=\p_{\mu}\xi+\gamma_{\mu}\varphi$ and $\delta \bar{A}_{\mu}=\p_{\mu}\bar{\xi}+\bar{\varphi}\gamma_{\mu}$ and a equivalent set for $B_{\mu}$. \footnote{ Since the gamma matrices take the gamma trace of the spinor-vector fields one could think that they are analogue to the Weyl tranformations of the spin-2 case. This becomes the analogies even more interesting.} 

Taking the variation of the simple sum of the actions we have explicitly:

\begin{align}
\delta \left(S^{(3)}_{+3/2}[A] + S^{(3)}_{-3/2}[B]\right) =& \int d^3x \left\lbrace\bar{f}^{\mu}(\Lambda)\left[\gamma_{\nu}\gamma_{\mu}f^{\nu}(\psi+ \alpha\chi) + \frac{1}{2}\Delta_{\alpha \mu \nu \beta}E^{\alpha \beta}f^{\nu}\left(\frac{\psi}{m_-} - \frac{\alpha \chi}{m_+}\right)\right]\nonumber\right.\\ 
+& \left. \left[\bar{f}^{\mu}(\psi+ \alpha\chi)\gamma_{\nu}\gamma_{\mu} + \frac{1}{2}\bar{f}^{\mu}\left(\frac{\psi}{m_-} - \frac{\alpha \chi}{m_+}\right)\Delta_{\alpha \mu \nu \beta}E^{\alpha \beta}\right]f^{\nu}(\Lambda)\right\rbrace,
\end{align}

\noindent which automatically defines the soldering currents  ${J}_{\mu}$ e $\bar{J}_{\mu}$ as the expressions between brackets multiplying $\bar{f}^{\mu}(\Lambda)$ and $f^{\mu}(\Lambda)$ respectively. On the definition of the soldering currents, which some times are called Nother currents we have given some more detailed discussion in \cite{parity} after expression (19); there the reader can see that there is some freedom in defining it. After all, we have the following result:

\begin{equation}
\delta\left(S^{(3)}_{+3/2}[A] + S^{(3)}_{-3/2}[B]\right) = \int d^3x\, \left[\bar{f}^{\mu}(\Lambda)J_{\mu} + \bar{J}_{\nu}f^{\nu}(\Lambda)\right]\label{noninv}
\end{equation}

\no Once the simple sum of the actions in (\ref{noninv}) is non invariant under the set of symmetries (\ref{sym1}) and (\ref{sym2}) becomes needed the introduction of auxiliary fields $\bar{H}_{\mu}$ and $H_{\mu}$ which are defined such as their transformations are given by $\bar{f}^{\mu}(\Lambda) = \delta\bar{H}^{\mu}$ e $f^{\nu}(\Lambda) = \delta H^{\nu}$. Then, after some rearrangements we have:

\begin{equation}
\delta \left[S^{(3)}_{+3/2}[A] + S^{(3)}_{-3/2}[B] - \int d^3x\,\,\left(\bar{H}^{\mu}J_{\mu} + \bar{J}_{\nu}H^{\nu}\right)\right]=  -\int d^3x\,\, \left(\bar{H}^{\mu}\delta J_{\mu} + \delta \bar{J}_{\nu}H^{\nu}\right),
\end{equation}

\no where the current variations are given by:

\begin{equation}
\delta J_{\mu} = (1+\alpha^2)\gamma_{\nu}\gamma_{\mu}f^{\nu}(\Lambda) + \frac{1}{2}\left(\frac{1}{m_-} - \frac{\alpha^2}{m_+}\right)\Delta_{\alpha \mu \nu \beta}E^{\alpha \beta}f^{\nu}(\Lambda)\label{dj},
\end{equation}

\begin{equation}
\delta\bar{J}_{\nu} = (1+\alpha^2)\bar{f}^{\mu}(\Lambda)\gamma_{\nu}\gamma_{\mu} + \frac{1}{2}\left(\frac{1}{m_-} - \frac{\alpha^2}{m_+}\right)\bar{f}^{\mu}(\Lambda)\Delta_{\alpha \mu \nu \beta}E^{\alpha \beta}\label{dj2},
\end{equation}

\noindent The last terms in (\ref{dj}) and (\ref{dj2}) differently of the first case do can be rewritten in terms of  $\delta H^{\nu}$ and $\delta\bar{H}^{\mu}$ respectively, but it would be more difficult to invert the $H$'s in terms of the $K$'s in the next step. Besides due to the presence of an additional derivative in these terms, the result of this inversion would be non-local and that is why we choose $\alpha^2 = {m_+}/{m_-}$, in order to eliminate these terms. In such a way we define the soldered action in terms of the auxiliary fields as:

\begin{equation}
S_S= S^{(3)}_{+3/2}[A] + S^{(3)}_{-3/2}[B] - \int d^3x (\bar{H}^{\mu}J_{\mu} + \bar{J}_{\nu}H^{\nu} - (1+\alpha^2)\bar{H}^{\mu}\gamma_{\nu}\gamma_{\mu}H^{\nu}), \label{ASOLDA}
\end{equation}

\no which is finally invariant under the set of symmetries (\ref{sym1}) and (\ref{sym2}). But the auxiliary fields must be eliminated of the final result, and then we use their equations of motion to obtain:

\begin{equation}
S_S^{(4)}= S^{(3)}_{+3/2}[A] + S^{(3)}_{-3/2}[B] + \frac{1}{2(1+\alpha^2)}\int d^3x (\bar{J}^{\mu}J_{\mu} - \bar{J}_{\mu}\gamma^{\nu}\gamma^{\mu}J_{\nu}).\label{jj2}
\end{equation}

Substituting $\bar{J}^{\mu}$ and  $J_{\mu} $ in (\ref{jj2})  and defining invariant soldering fields $\psi_{\mu} = \alpha A_{\mu} - B_{\mu}$ and $\bar{\psi}_{\mu}=\alpha \bar{A}_{\mu}-\bar{B}_{\mu}$ we end up with the final fourth order seldered action:

\begin{align}
S_S^{(4)} = &\frac{1}{2(1+\alpha^2)} \int d^3x \left[\bar{f}^{\mu}(\psi)\gamma_{\nu}\gamma_{\mu}f^{\nu}(\psi) + \frac{(m_+ - m_-)}{2(m_+m_-)}\bar{f}^{\mu}(\psi)\Delta_{\alpha \mu \nu \beta}E^{\alpha \beta}f^{\nu}(\psi)\right. \label{sterc111}\\\nonumber 
& \left. - \frac{1}{4(m_+m_-)}\bar{f}^{\mu}(\psi)E^{\alpha \beta}\gamma_{\alpha}\Delta_{\mu \theta \beta \tau}\gamma_{\omega}E^{\theta \omega}f^{\tau}(\psi)\right].
\end{align}\label{ult} 

There are several similarities of this result with the spin-2 case where we have found the linearized version of the New Massive Gravity in \cite{parity}. The second order term which would be the Eintein-Hilbert term has been appeared with wrong sign as in the spin-2 case. The generalized soldering procedure has produced a third order term which seems to be of the kind of the third order topological Chern-Simons term, such term gets rid of when we do $m_+=m_-$. The fourth order term is the same we have obtained through the alternative Nother gauge embedment approach in \cite{nges32}. We have interpreted it as a kind of $K$ term from the NMG model. Then if one choose equal masses we end up with a spin-3/2 model analogue to the NMG gravity theory at least in the linearized version. Despite the gauge invariance of the model, we also have some similarities since the fourth order model is invariant under reparametrizations of the type $\delta \psi_{\mu}=\p_{\mu}\xi$ and gamma Weyl-like transformations $\delta \psi_{\mu}=\gamma_{\mu}\varphi$.

\section{Conclusion}

The recent doublet models we have suggested in \cite{nges32} are recovered through the generalized soldering procedure. In particular we have observed that starting with the dublet model (\ref{dublet}) one can lower the order in derivatives by considering the introduction of an auxiliar field and then perform a rotation in the fields in order to decouple the lagrangian obtaining then a pair of first-order self-dual models with different helicities. The same happens with spin-1 and spin-2 theories. The procedure is also some times understood as to take the square root of the dublet model.

We have also demonstrated that the generalized soldering procedure of two second order self-dual models with different helicities $+3/2$ and $-3/2$ and also different masses $m_+$ and $m_-$ gives rise to a generalized version of (\ref{dublet}) where an interference term is present. Such terms are typical in the generalized procedure. Here it is a first order Chern-Simons like term which has a coefficient $(m_+-m_-)$. Then it turns out to zero if we choose equal masses.

By soldering two third-order self-dual models with different masses and helicities one can obtain the fourth order dublet model that we also have suggested in \cite{nges32}. However here we have an interference term of third order in derivatives, which generalizes the previous result. Such term is similar to the spin-2 topological Chern-Simons term and also disappear if we choose equal masses. 

In the case of spin-2 theories the enigmatic combination on the mass term of the symmetric rank-2 tensor $h_{\mu\nu}$ and its trace $h$ provide precisely the correct description of the spin-2 mode alone without any ghosts. It is interesting to notice that this mass term is automatically generated through the soldering procedure of two second order self-dual models. Here we have also generated the spin-3/2 mass term which is of the form $\epsilon^{\mu\nu\alpha}\bar{\psi}_{\mu}\gamma_{\nu}\psi_{\alpha}$. It is interesting to notice that in the spin-1 and in the spin-2 cases the soldering of two second order self-dual models in $D=2+1$ gives rise to the Proca and Fierz-Pauli dublet models respectively. Both of them are dimension independent. Here the mass term in (\ref{dublet}) is proportional to the Levi-Civita symbol, then restricted a priori to the three dimensional space. One can however rewrite it by  considering the identity $\epsilon^{\mu\nu\alpha}\gamma_{\alpha}=\gamma^{\mu}\gamma^{\nu}-\eta^{\mu\nu}$.

In our point of view, the results we have obtained here reinforces the dublet models we have suggested in \cite{nges32} since they are recovered by means of the soldering of singlet models as also happens in the spin-1 and spin-2 cases. Taking the spin-2 case for example, the second order dublet model (\ref{dublet}) is similar to the Fierz-Pauli theory, while the fourth order dublet model is analogous to the linearized version of the $NMG$ theory.

All along the work we have been used the word equivalent in order to relate the pair of self-dual models and the dublet models. Such equivalence must be understood in the same spirit of the work [11] where the interest is a connection between the theories in the lagrangian level. In fact in the soldering procedure there is a priori no guarantee of quantum equivalence between the pair of self-dual theories describing opposite helicities and the final soldered field theory. In order to stablish the equivalence one need to construct a master action interpolating between the pair of singlet models and the dublet models. This is precisely the aim of a work in progress where we are constructing master actions interpolating among the three self-dual models, between the two dublet models and among the pair of self-duals and the dublet models in sintony with [10] where we have done some thing similar with the spin-2 case. It would be very unusual if the existence of such ``equivalence" at the classical level do not persist at the quantum level.

\section{Acknowledgements}

E.L.M is supported by \textbf{CNPq} (449806/2014-6). Sá de L. D. is supported by CAPES. We thank Prof. Denis Dalmazi, for several discussions.

\end{document}